\documentclass[twocolumn,showpacs,letterpaper,preprintnumbers,amsmath,amssymb]{revtex4}
\usepackage{epsfig}
\usepackage{dcolumn}
\usepackage{bm}
\begin{document}


\title{Superconductivity and Field-Induced Magnetism in Pr$_{2-x}$Ce$_x$CuO$_4$ Single Crystals}
\author{J.E.~Sonier$^{1,5}$, K.F.~Poon$^{1}$, G.M.~Luke$^{2,5}$, P.~Kyriakou$^{2}$, R.I.~Miller$^{3,*}$, R.~Liang$^{3,5}$,
C.R.~Wiebe$^{2}$, P.~Fournier$^{4,5}$ and R.L.~Greene$^{6}$}
\affiliation{$^1$Department of Physics, Simon Fraser University, Burnaby, British Columbia V5A 1S6, Canada \\
$^2$Department of Physics \& Astronomy, McMaster University, Hamilton, Ontario, L8S 4M1, Canada \\
$^3$Department of Physics and Astronomy, University of British Columbia, Vancouver, British Columbia V6T 1Z1, Canada \\
$^4$D\'{e}partment de Physique, Universit\'{e} de Sherbrooke, Qu\'{e}bec J1K 2R1, Canada \\
$^5$Canadian Institute for Advanced Research, Toronto, Ontario, Canada \\
$^6$Center for Superconductivity Research, Department of Physics, University of Maryland, College Park, Maryland 20742, USA}
\date{\today}


\begin{abstract}
We report muon-spin rotation/relaxation ($\mu$SR) measurements on single crystals of the electron-doped high-$T_c$
superconductor Pr$_{2-x}$Ce$_x$CuO$_4$. In zero external magnetic field, superconductivity is found to coexist with Cu spins
that are static on the $\mu$SR time scale. In an applied field, we observe a $\mu^+$-Knight shift that is primarily due to
the magnetic moment induced on the Pr ions. Below the superconducting transition temperature $T_c$, an additional source of
static magnetic order appears throughout the sample. This finding is consistent with antiferromagnetic (AFM) ordering of the
Cu spins in the presence of vortices. We also find that the temperature dependence of the in-plane magnetic penetration
depth $\lambda_{ab}$ in the vortex state resembles that of the hole-doped cuprates at temperatures above $\sim 0.2$~$T_c$.
\end{abstract} 
\maketitle

While there now exists a large body of convincing experimental work on high-$T_c$ superconductors with hole-type carriers,
the intrinsic properties of the electron-doped cuprates $R_{2-x}$Ce$_x$CuO$_4$ ($R \equiv$~La, Pr, Nd, Sm or Eu) have
remained very elusive. Part of the problem stems from the difficulty of preparing single phase superconducting samples. An
additional complication is the presence of both Cu and rare-earth ($R$) moments. The interplay between these two magnetic
sublattices has been studied extensively in the undoped parent compounds $R_2$CuO$_4$ \cite{Sachi:97}. In Ce-doped samples,
there is the possibility that the magnetic exchange interactions involving the CuO$_2$ planes play an important role in
superconductivity. There is also some uncertainty about the symmetry of the superconducting order parameter. Although,
recent phase sensitive \cite{Tsuei:00} and angle-resolved photoemission \cite{Armitage:01} experiments are consistent with
$d$-wave pairing symmetry, measurements of thermodynamic quantities such as the in-plane magnetic penetration depth
$\lambda_{ab}$, generally do not show the expected linear temperature dependence at low $T$ that is characteristic of a
``clean'' superconductor. Many of the early results for $\lambda_{ab}(T)$ in electron-doped cuprates were obtained in the
Meissner phase of Nd$_{2-x}$Ce$_{x}$CuO$_4$ (NCCO) thin films and single crystals. While these results favored $s$-wave
symmetry \cite{Wu:93,Andreone:94,Anlage:94}, it has since been recognized that the paramagnetism of the Nd ions strongly
affects the low-$T$ behavior of $\lambda_{ab}(T)$ \cite{Cooper:96,Alff:99,Prozorov:00a,Kokales:00}. Consequently, attention
has shifted primarily to the Pr$_{2-x}$Ce$_x$CuO$_4$ (PCCO) system, where crystal electric field (CEF) splitting of the
Pr$^{3+}$ $J \! = \! 4$ manifold results in a nonmagnetic singlet ground state. Unfortunately, more recent measurements of
$\lambda_{ab}(T)$ in PCCO thin films and crystals using a variety of techniques have not yielded consistent results
\cite{Alff:99,Kokales:00,Prozorov:00a,Ku:01,Skinta:02a,Skinta:02b,Biswas:02}.

To avoid extrinsic effects related to the sample surface, we have utilized the $\mu$SR technique. $\mu$SR is a local probe
of the bulk and is one of the few techniques that can measure $\lambda_{ab}(T)$ in the vortex state --- sensitive to the
supercurrents circulating around the vortex cores, rather than the screening currents which flow near the sample surface.
Attempts to accurately measure $\lambda_{ab}(T)$ in NCCO by $\mu$SR have been precluded by the considerable Nd-moment
contribution to the muon-spin depolarization rate \cite{Luke:97}. In this Letter, we report on the first $\mu$SR study of
superconducting PCCO single crystals.

We studied three single crystals of PCCO grown by a directional solidification technique in Al$_2$O$_3$ crucibles
using a CuO-based flux \cite{Peng:91}. Reduction of oxygen to induce superconductivity was achieved by encapsulating each
single crystal in polycrystalline PCCO in the presence of flowing Ar at 900-1000 $^{\circ}$C (Ref.~\cite{Brinkmann:96}).
Here we report on representative data from one of the single crystals, noting that qualitatively similar results were
obtained in a partial $\mu$SR study of the other two. The crystal was 0.07~mm thick, having an $\hat{a}$-$\hat{b}$ plane
surface area of $\sim 6.5$~mm$^2$ and a mass of 3.74~mg. Although resistivity measurements indicate a $T_c$ value of
25(1)~K, bulk susceptibility $\chi^{\parallel}$ measurements show that the diamagnetic signal assumes a constant value below
$\sim \! 16$~K (see Fig.~1 inset).

The measurements were carried out on the M15 and M20B positive muon ($\mu^+$) beam lines at TRIUMF, Canada. In a $\mu$SR
experiment the spin of the implanted $\mu^+$ precesses at a frequency $\omega_{\mu} \! = \! \gamma_{\mu}B_{\mu}$, where
$B_{\mu}$ is the local field at the $\mu^+$ site and $\gamma_\mu \! = \! 13.5534$~kHz G$^{-1}$ is the muon gyromagnetic
ratio. The PCCO single crystals are the smallest samples studied to date by conventional $\mu$SR. Consequently, a special
arrangement of scintillator detectors inside a He gas-flow cryostat was needed to eliminate the contribution from muons that
did not stop in the sample. Unfortunately, this setup is not possible in a dilution refrigerator, so 2.3~K is the lowest
temperature that could be reached. Measurements were taken both in zero external field (ZF) and in a transverse field (TF)
geometry with $H$ directed perpendicular to the CuO$_2$ planes.

\begin{figure}
\centerline{\epsfxsize=3.2in\epsfbox{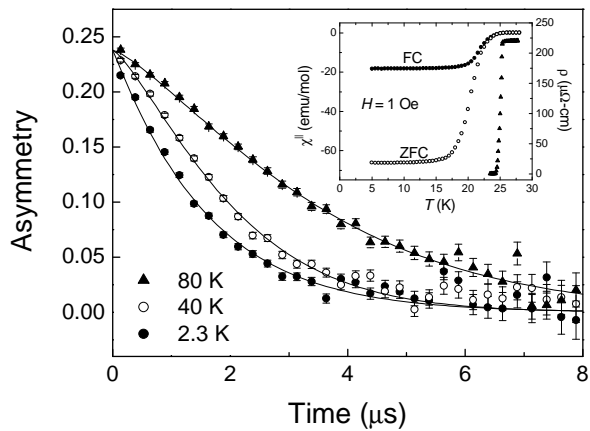}}
\caption{Time evolution of the asymmetry ({\it i.e.} $A(t) \! = \! A P(t)$, where $A$ is the signal amplitude) in zero
external field at $T \! = \! 80$~K, 40~K and 2.3~K. The initial muon-spin polarization was perpendicular to the
$\hat{c}$-axis. Inset: Temperature dependence of the resistivity (solid diamonds), and the bulk susceptibility at ${\bf H}
\! \parallel \! {\bf c} \! = \! 1$~Oe under FC and ZFC conditions.}
\end{figure}

In ZF, the muon-spin polarization $P(t)$ is relaxed by magnetic moments of electronic and nuclear origin. As shown in
Fig.~1, the relaxation rate of the ZF-$\mu$SR signal in PCCO increases with decreasing $T$. The absence of discrete
frequency oscillations indicates that there is no onset of spontaneous magnetic order at temperatures above 2.3~K.
Additional measurements carried out in a longitudinal-field geometry indicate that the muon spin is depolarized by randomly
oriented internal fields which are static on the $\mu$SR time scale. Since the Pr ions have a nonmagnetic singlet ground
state and the nuclear contribution to $P(t)$ is small, we attribute the source of the magnetism to disordered Cu spins.

TF-$\mu$SR and magnetization measurements were carried out under both field-cooled (FC) and zero-field cooled (ZFC)
conditions. Typical asymmetry spectra are shown in the inset of Fig.~2. In the ZFC case, pinning at the sample edges
prevents flux from entering the bulk at $T \! = \! 2.3$~K and $H \! < \! 300$~Oe. Thus, the ZFC time spectrum in Fig.~2
resembles that observed in ZF. We note that this is also the case in high-quality YBa$_2$Cu$_3$O$_{7-\delta}$ single
crystals. Above 300~Oe, flux fully penetrates the sample at $T \! = \! 2.3$~K, but pinning centers prevent the formation of
an equilibrium vortex lattice configuration. On the other hand, a well-ordered vortex lattice is achieved in the FC
procedure. Fast Fourier transforms (FFT) of the corresponding muon-spin precession signals at $H \!  = \! 91$~Oe are shown
in Fig.~2. The FFT provides an approximate picture of the internal magnetic field distribution $n(B)$ \cite{Sonier:00}.
Above $T_c$, the linewidth is due to the electronic and nuclear magnetic moments. Below $T_c$, the lineshape is further
broadened and becomes asymmetric as a result of the inhomogeneous field distribution created by a lattice of vortices.
Because the contribution of the magnetic moments to the width of $n(B)$ increases with increasing $H$, we have restricted
the current study to low magnetic fields.

\begin{figure}
\centerline{\epsfxsize=3.2in\epsfbox{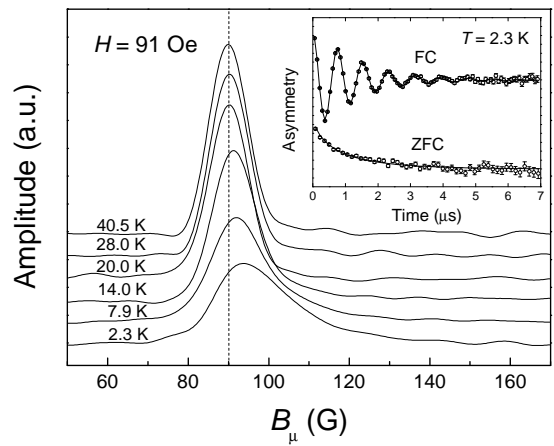}}
\caption{FFTs of the muon-spin precession signal in PCCO. Below $T_c$, the measurements were taken in an arbitrary sequence
as a function of $T$ under FC conditions. The dashed vertical line indicates the value of the external field $H \! = \!
91$~Oe. Inset: Asymmetry spectra taken at $T \! = \! 2.3$~K under both FC and ZFC conditions.}
\end{figure}

The stopping site of the $\mu^+$ has previously been identified in the related undoped parent compound Nd$_2$CuO$_4$ as near
an O atom midway between adjacent CuO$_2$ layers \cite{Luke:90}. Consistent with this site assignment, the TF-$\mu$SR signal
in PCCO shows a single well resolved signal with an average frequency shifted relative to the Larmor precession frequency of
$\mu^+$ in vacuum. The corresponding $\mu^+$-Knight shift is
\begin{equation}
K_{\mu}^{\parallel} =  \frac{B_0 - \mu_0 H}{\mu_0 H}
-4\pi \left( \frac{1}{3}-N \right) \rho_{\rm mol}\chi^{\parallel} \, ,
\label{eq:Knight}
\end{equation}
where $B_0$ is the average internal field sensed by the muons, the second term is a correction for Lorentz and bulk
demagnetization fields, $\rho_{\rm mol} \! = \! 0.01744$~mol/cm$^3$ is the molar density of PCCO, and $N \! \approx \! 1$ is
the demagnetization factor for a thin plate-like crystal. Figure~3 shows that $K_{\mu}^{\parallel}$ scales linearly with
$\chi^{\parallel}$ above $T_c$, indicating that the relative frequency shift is not induced by the $\mu^+$. In the normal
state there are several contributions to $K_{\mu}^{\parallel}$. Susceptibility measurements have established that the
field-induced moment on the Pr ions dominates the magnetic response of PCCO, which depends little on Ce-doping
\cite{Hundley:89}. Furthermore, as shown in the inset of Fig.~3, $\chi$ is highly anisotropic due to strong crystalline
electric-field (CEF) effects. For ${\bf H} \! \parallel \! {\bf c}$, there are smaller contributions to $\chi^{\parallel}$
from the tilting of the Cu moments out of the $\hat{a}$-$\hat{b}$ plane, Cu$^{2+}$-O$^{2-}$-Pr$^{3+}$ superexchange
interactions and from the polarization of the conduction electrons \cite{Foldeaki:94}. The Pauli paramagnetic susceptibility
$\chi_0$ of the conduction electrons gives rise to a $T$-independent Knight shift, $K_0 \! = \! A_0 \chi_0$, where $A_0$ is
the contact hyperfine coupling constant. The field induced moment on the Pr ions results in a $T$-dependent Knight shift
$K_f^{\parallel}$ consisting of two contributions: (i) the dipole-dipole interaction between the localized Pr $4f$-moments
and the $\mu^+$, and (ii) an indirect Rudermann-Kittel-Kasuya-Yosida (RKKY) interaction, producing a spin polarization of
the conduction electrons at the $\mu^+$ site. Assuming the Cu-moment contributions to be negligible, $K_{\mu}^{\parallel}$
at the axial symmetric $\mu^+$-site is given by
\begin{equation}
K_{\mu}^{\parallel} \approx K_0 + K_f^{\parallel} = K_0 + (A_{\rm c} + A_{\rm dip}^{zz}) \chi_f^{\parallel} \, ,
\label{eq:KfandK0}
\end{equation}
where $A_{{\rm c}}$ and $A_{\rm dip}^{zz}$ are the contact hyperfine and dipolar coupling constants, respectively, and
$\chi_f^{\parallel} \! \approx \! \chi^{\parallel} - \chi_0$. At the $\mu^+$ site we calculate $A_{\rm dip}^{zz} \! = \!
-663.6$~Oe/$\mu_B$. The solid line in Fig.~3 is a fit to Eq.~(\ref{eq:KfandK0}), yielding $A_{\rm c} \! = \!
3231$~Oe/$\mu_B$ and $K_0 \! > \! -4480$~ppm.

\begin{figure}
\centerline{\epsfxsize=3.2in\epsfbox{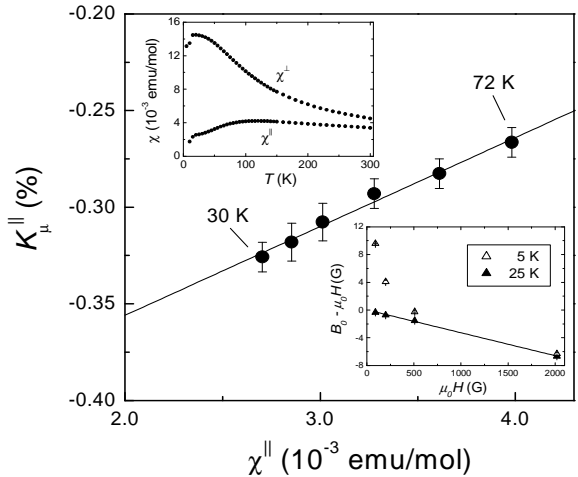}}
\caption{$K_{\mu}^{\parallel}$ vs. $\chi^{\parallel}$ at $H \! = \! 3$~kOe.
The solid line is a fit to Eq.~(\ref{eq:KfandK0}). Left inset: temperature dependence
of the bulk susceptibility at $H \! = \! 10$~kOe applied parallel
($\chi^{\parallel}$) and perpendicular ($\chi^{\perp}$) to the $c$-axis.
Right inset: magnetic field dependence of $B_0 - \mu_0 H$ at $T \! = \! 5$~K and
25~K. Each data point at $T \! = \! 5$~K represents a single FC measurement.}
\end{figure}

\begin{figure}
\centerline{\epsfxsize=3.2in\epsfbox{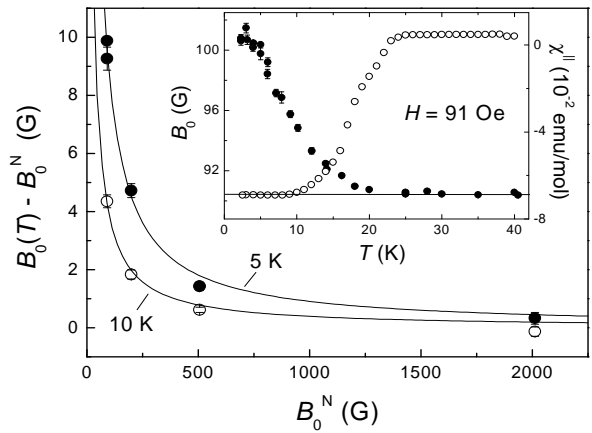}}
\caption{$B_0 (T) \! - \! B_0^{\rm N}$ vs. $B_0^{\rm N}$ at $T \! = \! 5$~K (solids circles)
and 10~K (open circles). The solid curves are described in the text.
Inset: Temperature dependence of $B_0$ and $\chi^{\parallel}$ taken under FC conditions.}
\end{figure}

Below $T_c$, there is a substantial increase of  $B_0$ that nearly coincides with the increase of the diamagnetic signal
under FC conditions (see Fig.~4). This cannot be explained by the reduction of $\chi_0$ that arises from the formation of
Cooper pairs. We first considered the possibility that the increased field is a manifestation of the so-called {\it
paramagnetic Meissner effect} \cite{Sigrist:95}, but the diamagnetic response observed in both the FC and ZFC magnetization
does not support this interpretation. A second possible origin, thought to be responsible for magnetorestriction
enhancements \cite{Zieglowski:88} and large diamagnetic shifts of $B_0$ \cite{Lichti:91} in the superconducting state of the
$R$Ba$_2$Cu$_3$O$_{7-\delta}$ system, is the $R$-moments induce screening currents in the CuO$_2$ layers. However, here the
Pr-moment is induced by the applied field, and thus is reduced below the normal-state value in regions well beyond the
vortex cores. This is inconsistent with the FFTs in Fig.~2, which imply that all muons stopping in the sample see an
increase in the local field. A third possible origin is that the vortices nucleate static Cu-spin order, as has been
observed in the superconducting state of underdoped La$_{2-x}$Sr$_x$CuO$_4$ \cite{Lake:02}. We note that while the
difference between $B_0$ in the superconducting and normal states is reduced with increasing $H$, this does not imply that
the additional source of field below $T_c$ weakens. For example, the solid curve in Fig.~4 is a fit assuming $B_0 (T) \! =
\! [ (B_0^{\rm N})^2 \! + \! (B_{\perp})^2]^{1/2}$, where $B_0^{\rm N}$ is the average internal field in the normal state
and $B_{\perp} \! \approx \! 43$~G and 28~G, at 5~K and 10~K, respectively. This indicates that the additional field
$B_{\perp}$ that appears at the $\mu^+$-site is primarily directed in the $\hat{a}$-$\hat{b}$ plane. Calculations show that
this is consistent with the La$_2$NiO$_4$ spin-structure that has been observed with $\mu$SR in the parent compound
Nd$_2$CuO$_4$ at low $T$ \cite{Luke:90}. Thus, long-range order of the Cu spins appears to be stabilized in PCCO by a low
density of vortices, perhaps due to the close proximity of the superconducting and AFM phases.

\begin{figure}
\centerline{\epsfxsize=3.2in\epsfbox{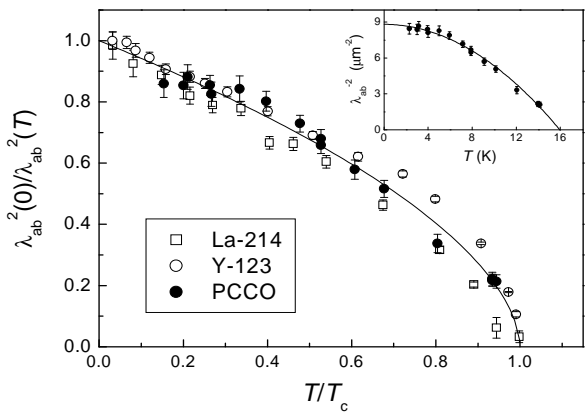}}
\caption{Normalized magnetic penetration depth $\lambda_{ab}^{2}(0)\lambda_{ab}^{2}(T)$ as a function of $T/T_c$ for single
crystals of PCCO at $H \! = \! 90$~Oe (solid circles), La$_{2-x}$Sr$_x$CuO$_4$ at $H \! = \! 2$~kOe (open squares)
\cite{Luke:97} and YBa$_2$Cu$_3$O$_{6.95}$ at $H \! = \! 5$~kOe (open circles) \cite{Sonier:99}. The solid curve is a guide
for the eye. Inset: $\lambda_{ab}^{-2}$ vs. $T$ in PCCO. The solid curve is a fit described in the text.}
\end{figure}

The in-plane magnetic penetration depth $\lambda_{ab}$ can be determined from the TF-$\mu$SR time spectra as described in
Ref.~\cite{Sonier:00}. In the normal state, the muon-spin precession signal is well described by the polarization function
$P(t) \! = \! e^{-(\sigma t)^K}\cos(\omega_{\mu}t + \phi)$, where $\beta \! \approx \! 1.6$ and $\sigma$ is the
temperature-dependent depolarization rate. The TF-$\mu$SR time spectra below $T_c$ were well fit to
\begin{equation}
P(t) = e^{-(\sigma_{\rm v} t)^{\beta}}\int n_{\rm v} (B) \cos(\gamma_{\mu} B t + \phi ) dB \, ,
\label{eq:polarization}
\end{equation}
where $n_{\rm v}(B)$ is the field distribution of an hexagonal vortex lattice derived from Ginzburg-Landau theory, and
$\sigma_{\rm v}$ is the depolarization rate due primarily to the effects of the magnetic moments --- which occur on a length
scale that is 10$^3$ times smaller than $\lambda_{ab}$. Fits at low $T$ which explicitly included a
variable transverse field component $B_{\perp}$, gave $\beta \! \approx \! 1.2$ and a diamagnetic shift for the
$c$-axis component of field. However, $\lambda_{ab}(T)$ could not be reliably obtained in this manner due to the decreased
sensitivity to $B_{\perp}$ with increasing $T$ and the competition with other fitting parameters.
Our analysis using Eq.~(\ref{eq:polarization}) shows that $\sigma_{\rm v}$ scales as $1/(T -\theta)$, where $\theta \! = \!
-18(5)$~K. At low $H$, only a small fraction of the implanted muons stop near the vortex cores.
As a result, the analysis is not very sensitive to the value of the coherence length $\xi_{ab}$. For example, increasing
$\xi_{ab}$ from 30~\AA~ to 60~\AA~ changes $\lambda_{ab}(0)$ by less than 2~\%. Figure~5 shows the temperature dependence of
$\lambda_{ab}^{-2}$. The absence of data below $T \! = \! 2.3$~K forbids an accurate determination of the limiting
temperature dependence of $\lambda_{ab}^{-2}(T)$. However, above this $\lambda_{ab}^{-2}(0)/\lambda_{ab}^{-2}(T)$ shows
reasonable agreement with that determined by $\mu$SR in the hole-doped systems La$_{2-x}$Sr$_x$CuO$_4$ and
YBa$_2$Cu$_3$O$_{7-\delta}$. The solid curve in the inset of Fig.~5 is a fit of the PCCO data to $\lambda_{ab}^{-2}(T) \! =
\! \lambda_{ab}^{-2}(0) [1-(T/T_c)^2]$, yielding $T_c \! = \! 15.9(2)$~K and $\lambda_{ab}(0) \! = \! 3369(73)$~\AA. The
values of $T_c$ and $\lambda_{ab}(0)$ indicate that the bulk of our sample is primarily underdoped. For example, Meissner
state measurements in Ref.~\cite{Skinta:02b} on an underdoped PCCO film reported $\lambda_{ab} \! = \! 3100$~\AA. This same
study showed evidence for a crossover from $s$-wave to $d$-wave behavior in the underdoped regime.

In conclusion, we have observed the onset of static magnetic order in the superconducting state of underdoped PCCO single
crystals with a field applied perpendicular to the CuO$_2$ plane. This result is consistent with vortex-induced AFM ordering
of the Cu spins throughout the sample. In addition, our measurements of $\lambda_{ab}^{-2}(T)$ in the vortex state appear
consistent with hole-doped cuprates down to $\sim 0.2$~$T_c$. However, a unique determination of the low-temperature
behavior by $\mu$SR awaits the availability of larger single crystals.

We are especially grateful to R.F. Kiefl for technical assistance and insightful discussions. This work was supported by the
Natural Sciences and Engineering Research Council of Canada, and in the United States by National Science Foundation Grant
DMR 01-02350.


\end{document}